# Formulation of Deformation Stress Fields and Constitutive Equations in Rational Mechanics

Xiao Jianhua

Measurement Institute, Henan Polytechnic University, Jiaozuo, China

**Abstract:** In continuum mechanics, stress concept plays an essential role. For complicated materials, different stress concepts are used with ambiguity or different understanding. Geometrically, a material element is expressed by a closed region with arbitral shape. The internal region is acted by distance dependent force (internal body force), while the surface is acted by surface force. Further more, the element as a whole is in a physical background (exterior region) which is determined by the continuum where the element is embedded (external body force). Physically, the total energy can be additively decomposed as three parts: internal region energy, surface energy, and the background energy. However, as forces, they cannot be added directly. After formulating the general forms of physical fields, the deformation tensor is introduced to formulate the force variations caused by deformation. As the force variation is expressed by the deformation tensor, the deformation stress concept is well formulated. Furthermore, as a natural result, the additive decomposition gives out the definition of static continuum, which determines the material parameters in constitutive equations. Through using the exterior differentials, the constitutive equations are formulated in general form. Throughout the paper, when it is suitable, the related results are simplified to classical results for easier understanding.

## Contents

## 1. Introduction

In classical elastic deformation of continuum mechanics, the stress concept is well defined based on experiments. However, when the stress concept is extended to complicated materials, the ambiguity becomes clear. Although the stress can be defined by formulating the deformation energy as the function of stress tensor contract with strain tensor in continuum mechanics [1-3], what is the engineering picture of stress still is a problem [4-5].

In engineering mechanics, each stress component $\sigma_{ij}(=\sigma_{ji})$ is well defined as the surface force acting on the surface $i$ on the direction $j$. The symmetry feature is defined as the static balance condition of element without rotation and is proved by the angular moment conservation. If we accept both, then 'the surface force acting on the surface $i$ on the direction $j$ is equal to the surface force acting on the surface $j$ on the direction $i$' will become a natural conclusion. Some logic ambiguity may reflect in his reasoning.

If one examines it by the tensor formulation rules [6-7] by above engineering definition, the stress should be a mixture tensor [8-9]. Why it is a covariant tensor in most conventional treatment?

In most researches, the stress tensor is formulated by force balance rules. The preexistence of force is assumed. The deformation is viewed as a coordinator transformation. Along this line, the stress tensor concept is studied again and again by researchers with different backgrounds.

In this paper, the geometrical field formulation will be studied.

In a resent paper [10], Walter Noll suggested a configuration formulation composed by an internal mapping and an external mapping. By his concept, the force system in configuration $\delta$ is a pair ($F_\delta^{\text{int}}, F_\delta^{ext}$), both force-balanced and torque-balanced. Here, the internal region mapping and external mapping are defined independently, but they are additive. So, the unit material element concept is replaced by a configuration unit. The continuum concept is roughly defined as the materially ordered sets. Frankly speaking, I cannot fully understand his mathematical treatment, especially the general treatment about deformation: it is a mapping.

Treating the deformation as a manifold mapping is very common [11-14]. In fact, it is accepted as a convention. The kernel in such a kind of treatment is that: taking deformation energy as the function of the deformation tensor, then the physical principles are applied to deduce the related equations. Once the stress tensor is established, the local differential coordinator transformation concept is used to formulate the required tensor form [12-14]. However, once the details for engineering applications are required, little can be obtained by engineers. So, an explicit formulation [15-18] is required.

In this research, the continuum concept is roughly defined as the materially ordered sets (as Walter Noll defined). Geometrically, a material element is expressed by a closed region with arbitral shape. The internal region is acted by distance dependent force (body force), while the surface is acted by surface force. Further more, the element as a whole is in a physical background which is determined by the continuum where the element is embedded. Physically, the total energy can be additively decomposed as three parts: internal region energy, surface energy, and the background energy. To formulate the physical energy quantities, the exterior differential tools are used, combining with the tensor formulation of deformation geometry in dragging coordinator system [8, 19].

By this formulation, the internal region energy is expressed by the mass center motion related energy. To express this energy (1-form), three covariant base vectors are required. The surface energy is related with the unit material surface configuration. Using the exterior differential tools, wedge product of two base vectors are used to describe the configuration surface (2-form). The background energy is related with the energy exchange between the unit material element with the medium as a global whole and it is expressed as the wedge product of three base vectors (3-form).

As an example [20], the potential force is related with 1-form (such as electrical field), the surface force is related with 2-form (such as magnetic field), the energy density is related the 3-form. Such a kind treatment is different from the treatment from the methods based on Lagrange or Hamilton quantity [21-25], although the implied philosophy is identical. Based on my personal feeling, although these researches are good enough in mathematic formulation, their physical reasoning is not so strong. It is true that the deformation is equivalent with a local differential coordinator transformation in mathematic treatment. However, what are their real physical pictures? What are the possible methods to do the engineering measurement? Or more directly, for engineers, how to understand the related formulation correctly? I do not think that these papers can answer these problems satisfactorily.

It is easy to introduce the mapping concept or its equivalent to formulate the deformation as a mapping in mathematic treatment. Even they are correct in general sense, only when the real physical laws are applied to mapping, the real physical process can be well expressed. In most researches, the physical laws are pre-determined. Hence, in essential sense, these researches put their main points on extending the known physical laws to general cases. So, for solving mathematic equations, they are power-full. Yes, it is a way to make scientific findings. However, for explaining the physical laws in essential sense, these methods are not so effective as they are expected. Then, a detailed and more engineering orientated treatment still is required. This is the general purpose of this paper.

Differing from above reasoning, this research will formulate the deformation concept by geometrical field in an explicit way. Here, the explicit means it is the real description of engineering operation way, rather than the mathematic operator ways. After that, based on the general energy form (decomposed as three items of exterior differentials), the stress concept will be introduced naturally through the deformation tensor.

After formulating the general form of energy by geometrical fields, the deformation tensor is introduced to formulate the energy variations caused by deformation. As the energy variation is expressed by the deformation tensor, the deformation stress concept is well formulated. Furthermore, as a natural result, the additive decomposition gives out the definition of 'static' continuum by zero deformation energy. Through using the exterior differentials, the motion equations are formulated in general form. Throughout the paper, when it is suitable, the related results are simplified to classical results.

## 2. Basic Notes on Related Formulation

To make the paper be easy understood, some concepts should be cleared. Furthermore, the mathematic formulation related with exterior differentials should be cleared.

### 2.1 Motion Concepts in Continuum Mechanics

The development of mater motion concept in physics is very important to understand the difficulty in understanding the stress concept in continuum mechanics.

There are two methods to describe matter motion. One is in Newton sense, by position displacement, the velocity is an example. Another is in Einstein sense, by gauge field variation.

Generally speaking, the modern researches put their efforts on establishing the position displacement interpretation, such as trace variation and motion path concept, or coordinator transformation to replace the gauge field variation. More abstract forms in mathematics are based on above two motion concepts in essential sense.

If displacement concept is used to describe motion, the force is defined to form the basic concept of energy. For simple motion, the energy variation is formulated as:

$$dE = \vec{f} \cdot d\vec{s} \qquad (1)$$

A vector space concept is formulated. The condition is that the environment has no other contribution.

However, as Walter Noll point-out, for the motion in continuum, as defined by A E Green, the energy variation is in the form:

$$dE = \sigma^i_j S^j_i \qquad (2)$$

The basic description of motion description causes different formulation system.

For simple motion in Newton particle mechanics, one has:

$$dE = \vec{f} \cdot d\vec{s} = m\frac{d^2\vec{s}}{dt^2} \cdot d(\frac{d\vec{s}}{dt}) = \frac{m}{2} d\left(\frac{d\vec{s}}{dt}\right)^2 \qquad (3)$$

The velocity plays the essential role. When the time parameter is implied, the potential force is defined as:

$$\vec{f} = -\nabla u(s) \qquad (4)$$

So, the result is:

$$dE = -\nabla u(s) \cdot d\vec{s} \qquad (5)$$

Hence, the potential function is named as energy function.

Combining both together, the Lagrange mechanics and Hamiltonian mechanics are formulated by defining the general energy function:

$$E = K \pm U \tag{6}$$

Its extremity defines the physical reality.

Summing above formulation, the position displacement dominate the motion concept.

However, the electromagnetic field gives out another kind of motion. The basic equations:

$$\vec{E} = -\nabla\phi + \mu\frac{\partial \vec{A}}{\partial t}, \quad \vec{B} = \nabla \times \vec{A} \tag{7}$$

They make the Newton motion concept becomes a non-complete concept.

The energy is in the form:

$$dE = \frac{1}{2}\mu B^2 + \frac{1}{2}\varepsilon E^2 \tag{8}$$

In fact, it is from the Maxwell equations, the relativity theory is established. The research on unifying the gravity force and the electromagnetic force originates from this basic fact.

My understanding is that: after the establishment of general relativity, the motion concept is dominated by different formulation on gauge variations.

In this research, the gauge variation is formulated in commoving dragging coordinator system by introducing the basic gauge vector transformation tensor. The Newton motion is also formulated correspondingly.

**2.2 Some notes about Tensor Theory in Continuum Mechanics**

In mathematics, the physical tensor is taken as the invariant objector. So, for different coordinator system selection, it is invariant in form. In most mathematic treatment, the $dx^i$ and $d\tilde{x}^i$ meet Riemannian geometrical invariant condition:

$$ds^2 = g_{ij}dx^i dx^j = \tilde{g}_{ij}d\tilde{x}^i d\tilde{x}^j \tag{9}$$

This equation means that when the coordinator system is transformed, the corresponding physical components are transformed in the coordinator forward transformation form (covariant component) or in the coordinator reverse transformation form (anti-covariant component).

In this research, this tensor formulation is applied to the initial dragging coordinator selection.

In deformation mechanics, the invariant of matter unit under discussion is achieved by given it a set of coordinators. The gauge tensor is used to describe the geometrical motion of the matter under consideration. For initial configuration, the differential distance vector is expressed as: $d\vec{s}_0 = \vec{g}_i^0 dx^i$. For current configuration, the differential distance vector is expressed as: $d\vec{s} = \vec{g}_i dx^i$. In traditional continuum mechanics, the deformation is expressed by the differential distance variation:

$$\delta s^2 = ds^2 - ds_0^2 = (g_{ij} - g_{ij}^0)dx^i dx^j. \tag{10}$$

Many formulations of strain tensor are based on this equation. This definition omits the local rotation, based on the false belief that: the local rotation is equivalent with rigid rotation, so has no contribution to distance variation. This concept is rejected in rational mechanics, where the deformation tensor is used to formulate the deformation motion.

In this research, the deformation is expressed as the vector differential [8-9]:

$$\delta \vec{s} = (\vec{g}_i - \vec{g}_i^{\,0})dx^i \tag{11}$$

So, the only way is to take the dragging coordinator system.

In treating fluid motion, the local rotation concept is used in continuum mechanics. Hence, the deformation tensor concept is used.

In this research, the deformation tensor $F_j^i$ is defined by:

$$\vec{g}_j = F_j^i \vec{g}_i^{\,0} \tag{12}$$

Therefore, the vector variation in deformation sense should be defined for a fixed point as:

$$\delta \vec{s} = (F_i^{\,j} - \delta_i^{\,j})\vec{g}_j^{\,0} \cdot dx^i \tag{13}$$

Where, the $dx^i$ is purely a scalar quantity.

Unfortunately, biased by the coordinator transformation concept, many researches define the deformation tensor as a coordinator transformation [26]. This is not sound in physical reasoning.

One way to make the deformation concept clear is to let:

$F_\alpha^i$ up-index defined on initial configuration, its summing is always about initial defined quantity; Low-index defined on current configuration, its summing is always about current defined quantity. As an example:

$$\vec{g}_\alpha = F_\alpha^i \vec{g}_i^{\,0}, \quad v^i \vec{g}_i^{\,0} = \tilde{v}^\alpha \vec{g}_\alpha = (\tilde{v}^\alpha F_\alpha^i)\vec{g}_i^{\,0}, \text{ so, } v^i = \tilde{v}^\alpha F_\alpha^i. \tag{14}$$

The main viewpoint is that: $dx^i$ are fixed quantity for the material unit under discussion.

Hence, the $v^i \vec{g}_i^{\,0} = \tilde{v}^\alpha \vec{g}_\alpha = (\tilde{v}^\alpha F_\alpha^i)\vec{g}_i^{\,0}$ only means an invariant physical field. As the physical field is changed after deformation, the differentials are:

$$\Delta v = \tilde{v}^\alpha \vec{g}_\alpha - v^i \vec{g}_i^{\,0} = (\tilde{v}^\alpha F_\alpha^i)\vec{g}_i^{\,0} - v^i \vec{g}_i^{\,0} \tag{15}$$

Therefore, the coordinator change has no relation with the deformation formulation. Generally speaking, for physical consideration, if the material has no intrinsic variation, the identity should be:

$$\Delta v = v^\alpha \vec{g}_\alpha - v^i \vec{g}_i^{\,0} = (v^\alpha F_\alpha^i)\vec{g}_i^{\,0} - v^i \vec{g}_i^{\,0} \tag{16}$$

All variations are caused by geometrical deformation. This is named as: geometrical deformation to distinguish the physical intrinsic variation (physical deformation).

**2.3 Notes on Exterior Differentials in Deformation Geometrical Formulation**

Generally speaking, in mathematic formulation of exterior differential theory, the vector component is determined by physical quantity (by differential operation on function defined on manifold). Hence, the coordinator differential is separated as quantity. Similarly, in deformation tensor formulation, the base vector is separated out also in a similar form.

Even so, the exterior differential tool [20, 27-28] cannot be applied without detailed analysis. Some form modifications are required.

The deep reasons are that: in physics, the physical field is produced by comparing the physical quantity difference (variation) between two near points. However, in deformation mechanics, the physical quantity difference (variation) at the same point is required.

For manifold U, functions $x^i$ on U (open set) are named as coordinates. Any functions on U can be written as: $f(x^1, x^2, ...., x^n)$. By exterior differential language, the local coordinator is interpreted as the coordinate 1-form: $dx^i$. For simple scalar potential field $\varphi$, the differential between two points (coordinator differences $dx^i$ play the vector role) is represented as a vector: $d\varphi = \frac{\partial \varphi}{\partial x^i} dx^i$. By deformation geometrical formulation adopted in this research, it should be represented as: $d\varphi = (\frac{\partial \varphi}{\partial x^i})(dx^{(i)} \vec{g}_i) = (\frac{\partial \varphi}{\partial x^i} dx^{(i)}) \cdot \vec{g}_i = \tilde{\varphi}^i \cdot \vec{g}_i$. Comparing both formulations, it is found that: in 1-form $d\varphi = \frac{\partial \varphi}{\partial x^i} dx^i$, the vector $dx^i \Rightarrow dx^{(i)} \vec{g}_i$ takes the role of covariant base vector multiplied by the coordinator increments; the $\tilde{\varphi}^i = (\frac{\partial \varphi}{\partial x^i} dx^{(i)})$ takes the role of anti-covariant vector component multiplied by the scalar component $\frac{\partial \varphi}{\partial x^{(i)}}$. To make useful simplification in deformation geometrical formulation, the scalar potential field differential should be understood as a vector in form:

$$d\varphi = (\frac{\partial \varphi}{\partial x^i} dx^{(i)}) \cdot \vec{g}_i = \tilde{\varphi}^i \cdot \vec{g}_i \tag{17}$$

Where (above and here after), the brackets is used to show the index in it is not viewed as a summing convention. That is to say, the physical vector component is decomposed as the multiplicative of quantity part and base vector part. The main point is that the $dx^i$ is not a vector in deformation theory (it is purely a scalar quantity).

Hence, in deformation formulation, the 1-form (like electrical field): $\omega = \omega_i dx^i$ in exterior differential is reformulated as:

$$\omega = \omega_{(i)} dx^i \vec{g}_i = \tilde{\omega}^i \vec{g}_i \tag{18}$$

The exterior differential 2-form (like magnetic field) is defined as: $\omega = \frac{1}{2} \omega_{ij} dx^i \wedge dx^j$. In deformation form adopted in this research, 2-form should be:

$$\omega = \frac{1}{2}[\omega_{(ij)} dx^i dx^j]\vec{g}_i \wedge \vec{g}_j = \frac{1}{2} \tilde{\omega}^{ij} \vec{g}_i \wedge \vec{g}_j \tag{19}$$

For the 3-form, $\omega = \frac{1}{3!} \omega_{ijk} dx^i \wedge dx^j \wedge dx^k$, similarly, it is reformulated as:

$$\omega = \frac{1}{3!}[\omega_{(ijk)} dx^i dx^j dx^k]\vec{g}_i \wedge \vec{g}_j \wedge \vec{g}_k = \frac{1}{3!} \tilde{\omega}^{ijk} \vec{g}_i \wedge \vec{g}_j \wedge \vec{g}_k \tag{20}$$

Using above formulations is based on the consideration that: for a fixed configuration, the exterior differential is used to define the physical field and their motion laws. As the physical fields ($\omega_i$, $\omega_{ij}$, and $\omega_{ijk}$) are well defined, by coordinating the material unit size in incremental coordinator form ($dx^i$), the fields ($\tilde{\omega}^i$, $\tilde{\omega}^{ij}$, and $\tilde{\omega}^{ijk}$) are well defined.

In deformation mechanics, the $dx^i$ (fixed on ordered materials which form continuum and is invariant) is viewed as anti-covariant coordinator selection, so the physical components is viewed as covariant form.

Summing above explanation, in deformation geometrical formulation of deformation mechanics, the exterior differential form is expressed as:

| | | |
|---|---|---|
| 0-form: | $\dfrac{f}{\sqrt{g}}$ | $\dfrac{f}{\sqrt{g_0}}$ |
| 1-form: | $\omega^i \vec{g}_i$ | $\omega^i \vec{g}_i^{\,0}$ |
| 2-form: | $\dfrac{1}{2}\omega^{ij} \vec{g}_i \wedge \vec{g}_j$ | $\dfrac{1}{2}\omega^{ij} \vec{g}_i^{\,0} \wedge \vec{g}_j^{\,0}$ |
| 3-form: | $\dfrac{1}{3!}\omega^{ijk} \vec{g}_i \wedge \vec{g}_j \wedge \vec{g}_k$ | $\dfrac{1}{3!}\omega^{ijk} \vec{g}_i^{\,0} \wedge \vec{g}_j^{\,0} \wedge \vec{g}_k^{\,0}$ |
| | In Current Configuration | In Initial Configuration |

The implied physical meaning is that: for a manifold, according to a given initial gauge selection, the exterior differentials are used to define the physical field.

As the deformation is introduced through deformation equation $\vec{g}_j = F_j^i \vec{g}_i^{\,0}$ (or in distance vector form $d\vec{s}_0 = \vec{g}_i^{\,0} dx^i$, and $d\vec{s} = \vec{g}_i dx^i$), so there is no need to change the exterior differentials system as they produce the required physical field, while the deformation the physical fields variation under the deformation.

**2.4 Notes on Material Invariance**

In classical deformation mechanics, the material feature is represented by elasticity parameters. On physical understanding, the material is defined by the elasticity parameters. In lattice dynamics theory, the elasticity parameters by physical field (quantum mechanics). So, it is clear, the so-called elasticity parameters are determined by physical fields. Based on this understanding, the physical fields ($\omega_i$, $\omega_{ij}$, and $\omega_{ijk}$) define the material features under deformation.

This viewpoint is buried in continuum mechanics. So, in this research, the original physical fields ($\omega_i$, $\omega_{ij}$, and $\omega_{ijk}$) are supposed as invariant. It is clear, for large deformation, they are not invariant. However, for incremental deformation, they are formulated as invariant in deformation mechanics formulation. This assumption is named as material invariance in this research.

**3. Physical Fields Variations Caused by Deformation**

Based on above discussions and formulation rules, the physical fields (defined by exterior differential form) variations caused by deformation are formulated in this sub-section.

In fact, for Gradient: $u = \frac{\partial u}{\partial x^i} \vec{g}_i^0$, the deformation caused variation is:

$$\delta u = \frac{\partial u}{\partial x^i}(F_i^{\ j} - \delta_i^{\ j})\vec{g}_j^0 = \delta u^j \vec{g}_j^0 \tag{21}$$

In deformation mechanics, the deformation caused fields (defined in exterior differential forms) variations are expressed as following.

**0-form:** $\quad \delta f = f\left(\frac{1}{\sqrt{g}} - \frac{1}{\sqrt{g_0}}\right) \tag{22}$

**1-form:** $\quad \delta\omega = \omega^i (F_i^{\ j} - \delta_i^{\ j})\vec{g}_j^0 = \delta\omega^j \vec{g}_j^0 \tag{23}$

**2-form:** $\quad \delta\omega = \frac{1}{2}\omega^{ij}(F_i^{\ k} F_j^{\ l} - \delta_i^{\ k}\delta_j^{\ l})\vec{g}_k^0 \wedge \vec{g}_l^0 = \frac{1}{2}\delta\omega^{kl}\vec{g}_k^0 \wedge \vec{g}_l^0 \tag{24}$

**3-form:** $\quad \delta\omega = \frac{1}{3!}\omega^{ijk}(F_i^{\ l} F_j^{\ m} F_k^{\ n} - \delta_i^{\ l}\delta_j^{\ m}\delta_k^{\ n})\vec{g}_l^0 \wedge \vec{g}_m^0 \wedge \vec{g}_n^0 = \frac{1}{3!}\delta\omega^{ijk}\vec{g}_i^0 \wedge \vec{g}_j^0 \wedge \vec{g}_k^0 \tag{25}$

Here (and after), the initial configuration is taken as reference.

For no deformation, all above quantities equal to zeros. By this formulation, one finds that:

$$\omega^i (F_i^{\ j} - \delta_i^{\ j})\vec{g}_j^0 \Rightarrow \omega_i dx^i \tag{26-1}$$

$$\frac{1}{2}\omega^{ij}(F_i^{\ k} F_j^{\ l} - \delta_i^{\ k}\delta_j^{\ l})\vec{g}_k^0 \wedge \vec{g}_l^0 \Rightarrow \frac{1}{2}\omega_{ij} dx^i \wedge dx^j \tag{26-2}$$

$$\frac{1}{3!}\omega^{ijk}(F_i^{\ l} F_j^{\ m} F_k^{\ n} - \delta_i^{\ l}\delta_j^{\ m}\delta_k^{\ n})\vec{g}_l^0 \wedge \vec{g}_m^0 \wedge \vec{g}_n^0 \Rightarrow \frac{1}{3!}\omega_{ijk} dx^i \wedge dx^j \wedge dx^k \tag{26-3}$$

Where, the left side quantities are the physical field variation, the right side quantities are the original physical field definition in exterior differential form. Detailed geometrical interpretation will be given in the next section.

## 4. Geometrical Formulation of Stress Fields

Geometrically, a material element is expressed by a closed region with arbitral shape. The internal region is acted by distance dependent force (body force), while the surface is acted by surface force. Further more, the element as a whole is in a physical background which is determined by the continuum where the element is embedded.

### 4.1. Deformation Description in Rational Mechanics

In classical mechanics, the unit material is under discussion. Its initial configuration can be expressed by three basic vectors. When the deformation happens, the three basic vectors are transformed into other forms. For different coordinator selections, the basic vectors will be transformed in the similar way. So, they are identical in mathematics sense.

However, for a material unit, the basic vectors ($dx^1, dx^2, dx^3$) or ($\vec{g}_1, \vec{g}_2, \vec{g}_3$) only defines the average shape. Or say, the length direction shape. If the material is not the cubic shape, however, it is arbitral shape, another set of basic vectors are needed ($\vec{g}^1, \vec{g}^2, \vec{g}^3$) (Referring Figure 1. Arbitral Shape of Material Unit). By Riemannian geometry tensor interpretation, they form surface base vectors. In exterior differential formulation, the 6 surfaces of the material unit is defined by $(dx^i \wedge dx^j)$ or $(\vec{g}_i \wedge \vec{g}_j)$. As an example, for a sphere shape, although the three rectangular direction length vectors forms the internal configuration, another three surface vectors are needed to forms the exterior surface configuration. In conventional geometry, the sphere surface is defined as: $(dx^1)^2 g_{11} + (dx^2)^2 g_{22} + (dx^3)^2 g_{33} = dr^2$.

Hence, in general, the complete deformation definition [9] is:

$$\vec{g}_i = F_i^j \vec{g}_j^0, \quad \vec{g}^i = G_j^i \vec{g}_0^j \tag{27}$$

Where, the 0 supper index or lower index is used for initial configuration.

When the selection $\vec{g}^i \cdot \vec{g}_j = \delta_j^i$ is made, the deformation can be simplified as: $\vec{g}_i = F_i^j \vec{g}_j^0$.

It is clear that: $G_i^l F_i^j = \delta^{lj}$. In this case, the volume invariant deformation is defined.

The physical reality of unit material is a finite element with finite volume. Hence, at least three basic vectors are required. In this research, they are ($\vec{g}_1, \vec{g}_2, \vec{g}_3$).

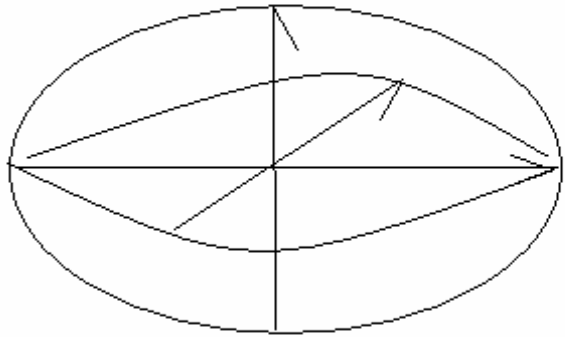

Figure 1. Arbitral Shape of Material Unit

Pure geometrically, a cubic element can be constructed by 9 independent points. One center, three covariant vectors passing through it, with 6 points; three surface passing the center point, each equipped with a normal vector (anti-covariant base vectors), 3 additional points are required. The center point can be constructed by the other 9 points, so it is not independent.

Hence, if and only if the 9 points are transformed in such a way that their relations are kept invariant, the coordinator transformations are acceptable.

However, in deformation mechanics, for arbitral shape element, another way should be found. This is described bellow.

### 4.1.1 Interior Line Description

For a material element center point ($x^1, x^2, x^3$), any line vector $d\vec{s}$ passing through the center point and take the intersection points with surface as its two ends is expressed by the covariant base vectors construction as: $d\vec{s} = dx^i \vec{g}_i$. By this formulation, the distance vector between two neighbor elements also is $d\vec{s} = dx^i \vec{g}_i$. Referring Figure 2. (Center Points Distance Descried as Interior Region Line).

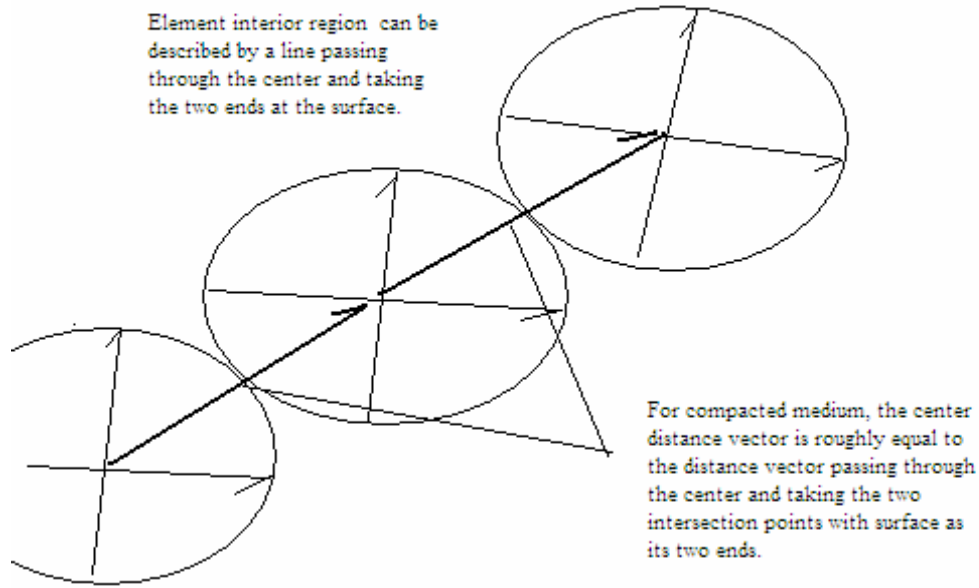

Figure 2. Center Points Distance Descried as Interior Region Line

By this definition, the material center point coordinator increments are ($dx^1, dx^2, dx^3$), and at the same time, they are the coordinator increments for any line passing through the center and connecting the two surface points.

After the whole medium is coordinated, the deformation is considered as the base vector variation. For initial referring configuration, $d\vec{s}_0 = dx^i \vec{g}_i^0$. After deformation, for current configuration, $d\vec{s} = dx^i \vec{g}_i$. Hence, the configuration variations are described.

The Riemannian geometry requires that: $ds^2$ is invariant to make the differential local coordinator transformation possible. However, for deformation, $ds^2$ is not invariant.

So, for deformation, the element configuration is defined by the numerical coordinator variation and the independent base vectors which are deformed. This is the essential meaning of deformation in rational mechanics.

Physically, this is the extension of point motion concept. Body force is based on this point of view. The displacement field $u^i$ (numerical) is based on this formulation system.

Chen's research [8-9] shows that:

$$F^i_j = \delta^i_j + \frac{\partial u^i}{\partial x^j} \tag{28}$$

This equation shows that the deformation tensor can be determined by measuring the displacement fields. For details, please see [8-9].

*4.1.2 Interior Surface Description*

On the other hand, the element exterior surface can be defined. The surface can be formulated as ($dx^1 dx^2, dx^2 dx^3, dx^3 dx^1$) in numerical sense, and in base vector sense as: ($\vec{g}^1, \vec{g}^2, \vec{g}^3$). (Referring Figure 3. Interior Surface Base Vectors).

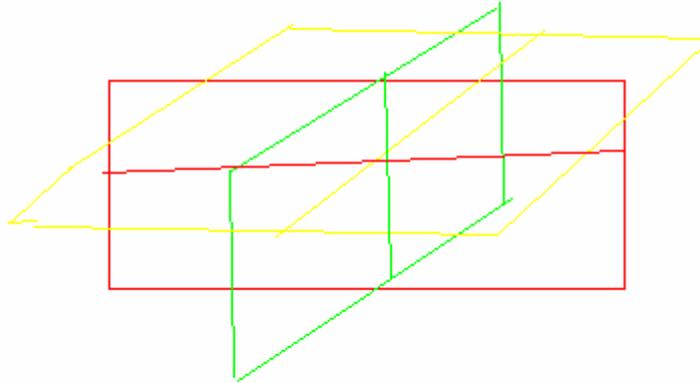

Figure 3. Interior Surface Base Vectors

The interior surfaces take the element center point as its geometrical center. For unit volume element, $\vec{g}^i \cdot \vec{g}_j = \delta^i_j$. This definition is used in continuum geometry to cover the whole medium. So, it is a global definition. Hence, it is named as interior surfaces. In geometrical theory, any surface passing through the element center can be decomposed as the linear combination of three basic surfaces. For this reason, the anti-covariant base vectors will not be used as the basic vectors for deformation.

In fact, each interior surface can be thought as a square which takes the four element center points on the square plane as the four corner points. Then, the surface is moved to take a material center as its geometrical center.

The stress concept in classical deformation mechanics is defined on these basic surfaces.

### *4.1.3 Exterior Surface Description*

The above interior surface definition forms the element as a cubic shape and its center is the material center. So, the real element surface is not described.

In this research, the region not closed in (by the real surface of element) is defined as exterior region.

So, here, the surface on element is defined as exterior surface. In this research, the exterior surfaces are expressed as: $dx^{(i)}dx^{(j)}\vec{g}_i \wedge \vec{g}_j$. The surface spanned out from the plane surface $dx^{(i)}dx^{(j)}\vec{g}_i \times \vec{g}_j$ (which centered at element center) is defined as exterior surface. Its geometrical meaning is explained in Figure 4. (Exterior Surface Definition).

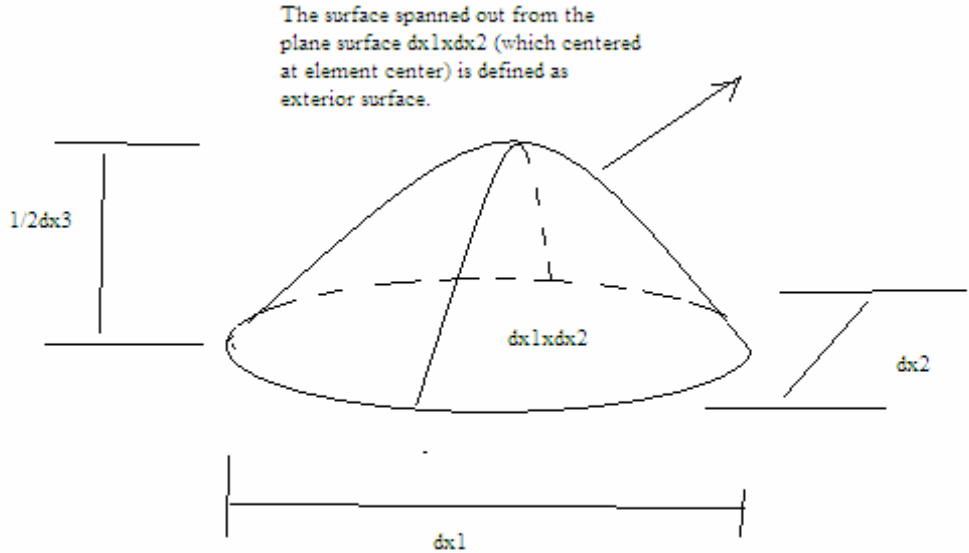

Figure 4. Exterior Surface Definition

Therefore, the interior surface passing material center $dx^1 dx^2 \vec{g}_1 \times \vec{g}_2$ forms the exterior surface spanned out on $\vec{g}_3$ direction based on right-hand chirality rules. The interior surface passing material center $dx^2 dx^1 \vec{g}_2 \times \vec{g}_1$ forms the exterior surface spanned out on $-\vec{g}_3$ direction. The two exterior surface forms a closed surface.

Similarly, other two forms can be explained. Hence, the general form $dx^{(i)}dx^{(j)}\vec{g}_i \wedge \vec{g}_j$ defines 6 spanned out surfaces. They form three kinds of configuration surfaces. As a geometrical convention, the spanned out surface should be the approximation of real surface shape of material element.

In deformation mechanics, the real area of spanned out surface is not a problem. This is because the area variation of spanned out surface is proportional with the area variation of interior surface.

For deformation $\vec{g}_i = F_i^j \vec{g}_j^0$, the exterior surface deformation is expressed as:

$$\vec{g}_i \wedge \vec{g}_j = F_i^k F_j^l \vec{g}_k^0 \wedge \vec{g}_l^0 \tag{29}$$

Therefore, the arbitral shape unit material deformation are fully described by the deformation tensor $F_i^j$. For simplicity, the exterior surface may be simply called as surface.

**4.2 Surface Force Derived Body Force**

In this research, the body force and surface force are acting on the material element.

The Body force is defined as: acting on everywhere. For deformation mechanics, the mass center or geometrical center point can be used to represents its action point.

The Surface force is defined as**:** acting on element surface. For deformation mechanics, there are 6 surfaces to form a closed region. Such a kind of closed region is defined as element unit.

In mechanics, the real problems are the body force will produce surface force and the vise versa. In conventional geometry, body forces are related with covariant base vectors related with point pair. Surface forces are related with anti-covariant base vectors related with surface.

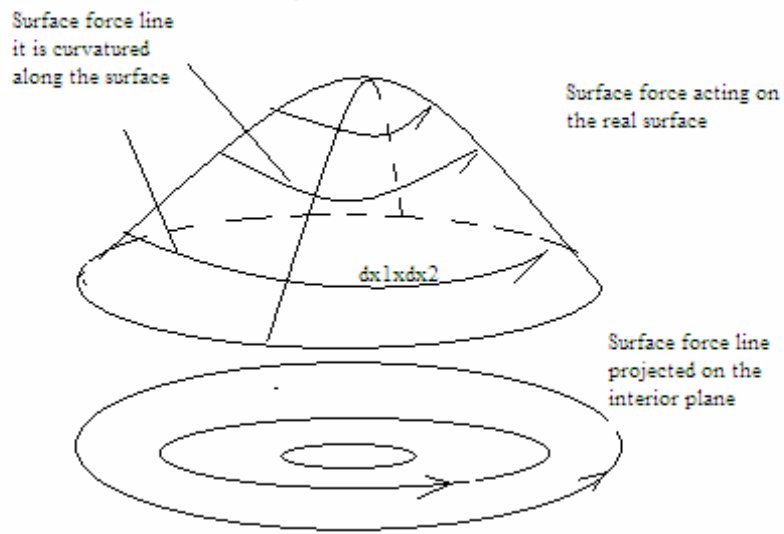

Figure. 5. 2-Form Surface Force

It is a fact that the surface force is explained by the external applied force. But, it cannot rule out the independent existence of surface force. For the independent existence of surface force, some papers are available. [29-35]

Here, as the exterior differential tools are used, the covariant base vectors are used and the anti-covariant base vectors are not used directly.

For a static unit element, in deformation mechanics, the pure 2-form surface force $\sigma^{ij}$ can be expressed as:

$$\sigma = \frac{1}{2}\sigma^{ij}\vec{g}_i \wedge \vec{g}_j \tag{30}$$

This surface force $\sigma^{ij}$ is acting on the surface $\vec{g}_i \wedge \vec{g}_j$ (right-hand chirality is selected). That means the force line is on the surface and takes the right-hand chirality as positive direction. For the surface geometrical meaning of surface force component $\sigma = \frac{1}{2}\sigma^{12}\vec{g}_1 \wedge \vec{g}_2$, referring Figure 5. (2-Form Surface Force).

Ii is clear that, for the upper-half surface $\vec{g}_1 \wedge \vec{g}_2$ and lower-half surface $\vec{g}_2 \wedge \vec{g}_1$ formed closed surface (tops separated by $dx^k$), the boundary of both is the edge of plane $dx^1 dx^2 \vec{g}_1 \times \vec{g}_2$. The force acting on the edge along the plane normal direction $\vec{g}_3$ can be viewed as the acting on the plane (or a body force distributed on the plane). This conclusion can be inferred from Stokes theorem. Here, a simple geometrical explanation is formulated.

$$\int_{surface} \frac{1}{2}\sigma^{12}\vec{g}_1 \wedge \vec{g}_2 = \frac{1}{2}\sigma^{12}\vec{g}_3 \text{, up-half surface} \tag{31-1}$$

$$\int_{surface} \frac{1}{2}\sigma^{21}\vec{g}_2 \wedge \vec{g}_1 = -\frac{1}{2}\sigma^{21}\vec{g}_3 \text{, low-half surface} \tag{31-2}$$

Hence, for the plane $dx^1 dx^2 \vec{g}_1 \times \vec{g}_2 = dx^1 dx^2 \vec{g}_3$, the body force derived from the closed surface force is:

$$\tilde{\sigma}^3 = \frac{1}{2}(\sigma^{12} - \sigma^{21}) \tag{31-3}$$

On the interior surface normal direction $\vec{g}_k = e_{ijk}\vec{g}_i \times \vec{g}_j$ (this operation is caused star operation in exterior differentials), the plane surface force can be formulated as:

$$\sigma = \frac{1}{2}(e_{ijk}\sigma^{ij})\vec{g}_k = \tilde{\sigma}^k \vec{g}_k \tag{32}$$

where, $\tilde{\sigma}^1 = \frac{1}{2}(\sigma^{23} - \sigma^{32})$, $\tilde{\sigma}^2 = \frac{1}{2}(\sigma^{31} - \sigma^{13})$, $\tilde{\sigma}^3 = \frac{1}{2}(\sigma^{12} - \sigma^{21})$ are element surface force derived body force components. It is clear that, the symmetry stress $\sigma^{ij} = \sigma^{ji}$ will cause zero normal direction body force $\tilde{\sigma}^i$. In exterior differential formulation, the symmetry parts are omitted. However, for deformation mechanics, for generality, the symmetry parts are reserved.

*This derived body force is acting on the internal region of the unit material element. Hereafter, it is named as interior body force.*

Through above formulation, it is concluded that: the symmetry stress $\sigma^{ij}$ can 'live' on a closed surface independently without producing net interior body force on normal direction.

Therefore, the surface forces are classified into two types: one kind ($\sigma^{ij}$) is acting on the exterior surfaces $dx^{(i)} dx^{(j)} \vec{g}_i \wedge \vec{g}_j$, that is they are defined on the 6 surfaces of unit element;

another kind is the net surface force acting on the interior surface (with area $dx^{(i)}dx^{(j)}$) normal direction. In essential sense, it is a body force distributed on the surface. In classical deformation mechanics, only the later is defined. In traditional elasticity, the stress is defined as the total contact force acting on the surface area.

The Equation (32) shows that **only the anti-symmetrical exterior surface stress components can be changed into interior body force.**

It exposes that: the stress symmetry denies the exterior surface force will produce internal body force. It is the static balance condition in traditional mechanics theory. The well-known interpretation is that: if the stress is not symmetrical, the element will rotate.

In fact, in traditional stress definition, the pure body force $\sigma^{(k)}$ acting in unique direction (here is taken as $k$) can be expressed as:

$$\vec{\sigma} = \sigma^{(k)} \vec{g}_k \quad \text{(no summation)} \tag{33-1}$$

One way is to define: $\vec{g}_k = \alpha_{kl} \vec{g}^l$, then one has:

$$\vec{\sigma} = \sigma^{(k)} \vec{g}_k = (\sigma^{(k)} \alpha_{kl}) \vec{g}^l = \sigma_{kl} \vec{g}^l \tag{33-2}$$

By this way, the body force is transformed into (surface force decomposition) classical stress $\sigma_{kl}$. This way is widely accepted in classical mechanics. The quantity $(\sigma^{(k)} \alpha_{kl})$ is defined as stress $\sigma_{kl}$. In fact, interpreting the $k$ index as the body force direction and the $l$ index as the surface direction, the stress $\sigma_{kl}$ is defined. This stress explanation (definition) is widely used in textbooks.

The above formulation also shows that, by suitable selecting the coordinator system, the classical stress always can be expressed as the pure normal force form: $\sigma = \sigma^{(k)} \vec{g}_k = \sigma^{(k)} \vec{g}^k$. This fact is well explained in mechanics textbook.

Reasoning from above analysis, the virtual surface force $\sigma = \frac{1}{2} \sigma^{ij} \vec{g}_i \wedge \vec{g}_j$ is ruled out in traditional stress concept.

**4.3 Body Force Derived Surface Force**

The traditional stress definition can be formulated by the 2-form formulation strictly.

In classical continuum mechanics, the body force is externally applied on the surface of material element, let $\vec{g}_k = e_{ijk} \vec{g}_i \wedge \vec{g}_j$ (star operation in exterior differentials) be the surface normal direction, one has:

$$\sigma = \sigma^k \vec{g}_k \tag{34-1}$$

By exterior differential rules, one has:

$$d\sigma = \frac{1}{2} \sigma^{ij} \vec{g}_i \wedge \vec{g}_j \tag{34-2}$$

Where,

$$\frac{1}{2} \sigma^{ij} = \frac{\partial \sigma^j}{\partial x^i} \tag{34-3}$$

Hence, the body force spatial variation will produce exterior surface forces.

In tensor geometrical formulation, one has:

$$\nabla \sigma^j = \frac{\partial \sigma^j}{\partial x^i} \vec{g}_i \qquad (35)$$

Hence, in traditional deformation mechanics, for cubic unit volume element, the classical stress tensor components $\sigma_{ij}^{cla}$ are defined as:

$$\sigma_{ij}^{cls} = \frac{1}{2}(\frac{\partial \sigma^j}{\partial x^i} + \frac{\partial \sigma^i}{\partial x^j}) = \frac{1}{2}(\sigma^{ij} + \sigma^{ji}) \qquad (36)$$

It shows that, the pure externally applied body forces variation can be decomposed as the average stress (interior surface force) on the interior surfaces of material element and the ideal 2-form exterior surface force. Therefore, external body force will produce average stress concept acting on the surface of material element, which will require that the stress is symmetry.

Hereafter, the symmetry parts of 2-form exterior surface stress will be named as classical stress, the anti-symmetrical parts will be named as rotational stress.

### 4.4 Intrinsic Stress Definition

Based on the analysis about surface derived body force and the body force derived surface force, it is shows that: 1) only the anti-symmetrical exterior surface force will produce interior body force, while the symmetry exterior surface force will not produce interior body forces; 2) the pure exterior body forces variation can be decomposed as the average surface force acting on the surface of material element (named as classical stress) and the difference surface force acting on the surface of material element (named as rotational stress).

Hence, the classical stress has no effects on the interior motion of material element. The interior region is separated out by physical features. This results support the two regions separated by closed surface adopted in this research about material element.

After separating the surface force derived interior body force from the exterior body force and separating the external body derived surface exterior stress from the virtual surface force, the unit material element force system is described by three sets: 1) the interior body force, which is acting on the internal region of the material element and taking the geometrical center of configuration as its center; 2) the exterior body force, which is applied to the material element from its surroundings; 3) the interior surface force, which is interacting with the internal region; 4) the exterior surface force, which is interacting with external region.

Now, we are ready to adopt the classical stress concept, which is defined as the body force acting on unit surface area on the surface normal direction.

The intrinsic stress is the stress not produced by the macro deformation. This is to distinguish the deformation stress purely caused by deformation.

#### *4.4.1 Internal Region Intrinsic Body Force and Self-spin*

For physical reality, the interior region can exist a virtual body force $f_{int}^i \vec{g}_i$, they will act on the internal side of the configuration surface. The general force in the internal region will be:

$$f = f_{int}^i \vec{g}_i \oplus \frac{1}{2}\sigma^{ij}\vec{g}_i \wedge \vec{g}_j \qquad (37)$$

Here, the symbol $\oplus$ is used to show that: the two vectors have different features, they cannot be added directly [20].

Note that, as $\vec{g}_i \wedge \vec{g}_j = -\vec{g}_j \wedge \vec{g}_i$, although for standard rectangular system (or initial configuration) if $\sigma^{ij} = \sigma^{ji}$, then $\frac{1}{2}\sigma^{ij}\vec{g}_i \wedge \vec{g}_j = 0$. But, for arbitral shape configuration, the symmetry may be broken. So, in general sense, $\sigma^{ij} \neq \sigma^{ji}$. Here, this point is different from pure exterior differentials.

By previous analysis (Equation (31)), on the material element interior surface, the exterior surface force will produce an internal surface force. If the interior body force components are thought as acting on the material center point, then, as the three internal surfaces passing through the element center, three principle intrinsic body force components can be defined as:

$$\sigma^1_{int} = \frac{1}{2}(\sigma^{23} - \sigma^{32}) + f^1_{int} \tag{38-1}$$

$$\sigma^2_{int} = \frac{1}{2}(\sigma^{31} - \sigma^{13}) + f^2_{int} \tag{38-2}$$

$$\sigma^3_{int} = \frac{1}{2}(\sigma^{12} - \sigma^{21}) + f^3_{int} \tag{38-3}$$

This body force is defined on the center point of material element.

It shows that the exterior surface force derived interior body force can affect the motion of interior region center point. They form a complete force system for center motion description.

This force will make the center point motion toward the force vector direction. Based on this point, the intrinsic definition about macro configuration static continuum is: $\sigma^i_{int} = 0$.

For macro configuration static continuum, the condition is expressed as:

$$\frac{1}{2}(\sigma^{23} - \sigma^{32}) + f^1_{int} = 0 \tag{39-1}$$

$$\frac{1}{2}(\sigma^{31} - \sigma^{13}) + f^2_{int} = 0 \tag{39-2}$$

$$\frac{1}{2}(\sigma^{12} - \sigma^{21}) + f^3_{int} = 0 \tag{39-3}$$

Considering an atom model, the $f^i_{int}$ is Coulomb electrical force towards the center point (so, they are negative). To balance this force, a rotational stress must exist on the atom shells. If the internal body force is negative (positive), the rotational stress is right-hand chirality (left-hand chirality).

So, generally, for macro configuration static continuum, the material element has self-spin. As the related detailed discussion may be to broad, here this topic will not be discussed further.

Simply, the above Equation (39) can be viewed as the definition of element unit self-spin.

### 4.4.2 External Region and Intrinsic Stress

The exterior region can exist a virtual body force $f^i_{ext}\vec{g}_i$, they will act on the exterior region of the configuration surface. The general force in the external region will be:

$$f = f^i_{ext}\vec{g}_i \oplus \frac{1}{2}\sigma^{ij}\vec{g}_i \wedge \vec{g}_j \tag{40}$$

The exterior force defined intrinsic stress components are:

$$\sigma_{11}^{cla} = \frac{\partial f_{ext}^1}{\partial x^1}, \quad \sigma_{22}^{cla} = \frac{\partial f_{ext}^2}{\partial x^2}, \quad \sigma_{33}^{cla} = \frac{\partial f_{ext}^1}{\partial x^1} \tag{41-1}$$

$$\sigma_{ij}^{cls} = \frac{1}{2}(\frac{\partial f_{ext}^j}{\partial x^i} + \frac{\partial f_{ext}^i}{\partial x^j}) + \frac{1}{2}(\sigma^{ij} + \sigma^{ji}), \quad i \neq j \tag{41-2}$$

For macro configuration static continuum, all classical intrinsic stress components are zero. So, the macro static conditions are:

$$\frac{\partial f_{ext}^{(i)}}{\partial x^i} = 0, \text{ exterior region} \tag{42-1}$$

and the static exterior surface is defined by equation:

$$\frac{1}{2}(\frac{\partial f_{ext}^j}{\partial x^i} + \frac{\partial f_{ext}^i}{\partial x^j}) + \frac{1}{2}(\sigma^{ij} + \sigma^{ji}) = 0, \quad i \neq j, \text{ on exterior surface} \tag{42-2}$$

One simple solution is:

$$f_{ext}^i = Const, \quad \sigma^{ij} = -\sigma^{ji} \text{ (on surface)} \tag{43}$$

That is: external body force is constant for everywhere, hence will not produce classical stress (so no macro deformation). At the same time, the exterior surface only has pure rotational stress. Hence, the material element can have self-spin. Referring Figure 6. (Simple Static Continuum Definition). Comparing with the geoid surface defined on the Earth, one may get better understanding about this concept.

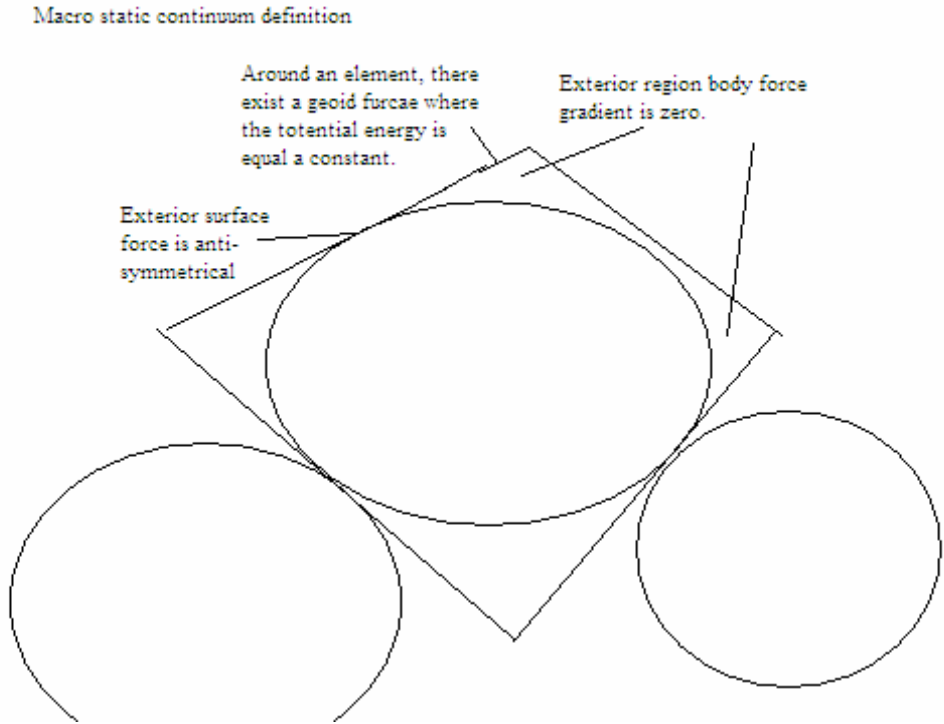

Figure 6. Simple Static Continuum Definition

### 4.4.3 Exterior Surface Configuration and Intrinsic Viscous Stress

In general cases, for a given exterior surface force $\frac{1}{2}\sigma^{ij}\vec{g}_i \wedge \vec{g}_j$, the static intrinsic viscous stress on surface can be defined as:

$$\eta_{ij}^{vis} = \frac{1}{2}(\frac{\partial f_{ext}^j}{\partial x^i} + \frac{\partial f_{ext}^i}{\partial x^j}) = -\frac{1}{2}(\sigma^{ij} + \sigma^{ji}), \quad i \neq j \tag{44}$$

The non zero viscous stress will cause the torsion of the exterior surface of material element without producing center points motion. Gel is a typical material in this static state, while the interfacial micro flows do exist.

The equation also shows that, if the surface intrinsic viscous stress is zero, the exterior body force must be a vector potential field (like magnetic field). That is: if $\eta_{ij}^{vis} = 0$, then

$$\frac{\partial f_{ext}^j}{\partial x^i} + \frac{\partial f_{ext}^i}{\partial x^j} = 0 \tag{45}$$

As in this case $\sigma^{ij} = -\sigma^{ji}$, the exterior surface force is purely rotational.

### *4.4.4 Macro Static Continuum Definition*

It is well known that the macro static continuum has micro dynamic motion. Based on above formulations, the macro static continuum is defined as:

$$\frac{1}{2}(\sigma^{ij} - \sigma^{ji}) + e^{ijk} f_{int}^k = 0, \text{ interior region} \tag{46-1}$$

$$\frac{\partial f_{ext}^{(i)}}{\partial x^i} = 0, \text{ exterior region} \tag{46-2}$$

$$\frac{1}{2}(\frac{\partial f_{ext}^j}{\partial x^i} + \frac{\partial f_{ext}^i}{\partial x^j}) + \frac{1}{2}(\sigma^{ij} + \sigma^{ji}) = 0, \quad i \neq j, \text{ on surface} \tag{46-3}$$

Hence, for given body forces (internal and external) and surface forces, the surface configuration is defined by equation:

$$\sigma^{ij} = -e^{ijk} f_{int}^k - \frac{1}{2}(\frac{\partial f_{ext}^j}{\partial x^i} + \frac{\partial f_{ext}^i}{\partial x^j}), \quad i \neq j \tag{47}$$

For a material element, all force components are the function of position referring to the element center. So, its solution gives out a closed surface. In theoretic sense, multiple solutions may exist. In this case, the element can has multi-possible configurations. In some cases, this phenomenon can be explained as the internal multi-scale features of materials.

It means that, for static state, the closed surface of material element has fixed configuration. The surface force on the closed surface is intrinsic. As a macro mechanics theory, the material element is viewed as a pure surface configuration as a whole.

Geometrically, the interior body force and exterior body force will force the material element produce self-rotation about its center. It is named as self-spin or spin of material element.

The material element surface has shedding function through chirality selection. This is very important to understand the complicated features of modem materials, especially artificial materials.

On physical sense, the body force is acting on everywhere and is related with the mass center motion, hence, the classical mechanics only consider the body force related with mass center motion. Its success on displacement concept is based on this intrinsic feature.

It is clear, when $\frac{1}{2}\sigma^{ij}\vec{g}_i \wedge \vec{g}_j = 0$, the material element has no chirality. Reasoning from this point, the classical mechanics only considers the electronic static potential field. The lattice dynamics takes this as its basic feature.

## 5. Constitutive Equation Formulation on Physical Bases

In electromagnetic field, the electric field is 1-form vector and the magnetic field is 2-form vector. Although they are different, their can act on the material element at the same time.

In fact, in physics theory, it is well accepted that an arbitral force can be decomposed as the additive of two parts: one is curl less (that is 1-form here), another one is divergent less (that is the 2-form).

Further more, the ideal element is the representative of average material features in continuum. Generally speaking, the deformation (or more exactly, the incremental deformation) is mainly produced by the exterior forces acting on the material element as a whole. So, physically, the stress should be defined through the general exterior force variation. This is the main task for this subsection.

Summing up above results, for a material unit, the general force is composed by original (virtual) body force and original (virtual) surface force in initial configuration:

$$f = f^i \vec{g}_i^0 \oplus \frac{1}{2}\sigma^{ij}\vec{g}_i^0 \wedge \vec{g}_j^0 \qquad (48)$$

After deformation, in current configuration, if the intrinsic features of materials have no variation (as it is supposed in infinitesimal deformation), the force is:

$$f = f^i \vec{g}_i \oplus \frac{1}{2}\sigma^{ij}\vec{g}_i \wedge \vec{g}_j \qquad (49)$$

By this way, the deformation $\vec{g}_i = F_i^j \vec{g}_j^0$ will produce the deformation force concept defined as:

$$\Delta f = f^i (F_i^j - \delta_i^j)\vec{g}_j^0 \oplus \frac{1}{2}\sigma^{ij}(F_i^k F_j^l - \delta_i^k \delta_j^l)\vec{g}_k^0 \wedge \vec{g}_l^0 \qquad (50)$$

Hence, referring to the initial configuration, without intrinsic variation of material features, (that is for invariant $f^i$ and $\sigma^{ij}$), the force produced by deformation is expressed as:

$$\tilde{f}^j = f^i (F_i^j - \delta_i^j), \text{ spring-like item} \qquad (51\text{-}1)$$

$$\tilde{\sigma}^{kl} = \sigma^{ij}(F_i^k F_j^l - \delta_i^k \delta_j^l), \quad k \neq l, \quad i \neq j, \text{ shear item} \qquad (51\text{-}2)$$

They give out the intrinsic definition of deformation force fields on physical bases. Here, the up-index is used for intrinsic force component to show the initial configuration is taken as the reference.

For exterior region, referring the intrinsic stress definition Equation (41), *the exterior deformation stress components* are defined as:

$$\sigma_1^1 = f_{ext}^1 \cdot (F_1^1 - 1), \quad \sigma_2^2 = f_{ext}^2 \cdot (F_2^2 - 1), \quad \sigma_3^3 = f_{ext}^3 \cdot (F_3^3 - 1) \qquad (52\text{-}1)$$

$$\sigma_j^i = f_{ext}^{(f)} F_j^i + \frac{1}{2}(\tilde{\sigma}^{ij} + \tilde{\sigma}^{ji}), \quad i \neq j \qquad (52\text{-}2)$$

***The above equation is the constitutive equation for general materials in deformation mechanics.*** (Hereafter, the power index ext will be dropt out).

If $\sigma^{ij} = -\sigma^{ji}$, then $\tilde{\sigma}^{ij} = -\tilde{\sigma}^{ji}$, the exterior surface deformation contribution is zero. The stress components in exterior region are purely produced by external body force. The element self-spin has no contribution to deformation stress.

For interior region, ***the interior deformation stress principle components are defined as:***

$$\tilde{\sigma}_{int}^{i} = \frac{1}{2} e^{ijk} (\tilde{\sigma}^{jk} - \tilde{\sigma}^{kj}) = \frac{1}{2} e^{ijk} \sigma^{mn} (F_m^j F_n^k - F_m^k F_n^j) \tag{53}$$

If $\sigma^{mn} = \sigma^{nm} = \sigma \delta^{mn}$, the surface deformation contribution is zero. This stress will be sensed by the material center point. If it is not zero, the center point will move its position. As a result, the surface configuration or exterior surface stress will be changed correspondingly. This topic is too broad to be included in this paper.

As the exterior deformation stress is taken as a convention in classical stress, the following formulation is named as the *Classical Formulation of Stress on Physical Bases*.

### *5.1 Isotropic Simple Solid Material*

In classical infinitesimal deformation theory, the classical strain is defined as:

$$\varepsilon_{ij} = \frac{1}{2}(F_j^i + F_i^j) - \delta_{ij} \tag{54}$$

Omitting higher order infinitesimals, one has:

$$\delta^{kl}(F_k^i F_l^j - \delta_k^i \delta_l^j) \approx 2\varepsilon_{ij} \tag{55}$$

For isotropic simple solid materials, defined as $f^1 = f^2 = f^3 = f^0$, $\sigma^{ij} = \sigma^{ji} = -\sigma^0, i \neq j$, (referring Figure 7. Isotropic Simple Solid Materials), the deformation force field is:

$$\tilde{f}^1 = f^i F_i^1 - f^1 = f^0(\varepsilon_{11} + \varepsilon_{12} + \varepsilon_{13}) \tag{56-1}$$

$$\tilde{f}^2 = f^i F_i^2 - f^2 = f^0(\varepsilon_{22} + \varepsilon_{21} + \varepsilon_{23}) \tag{56-2}$$

$$\tilde{f}^3 = f^i F_i^3 - f^3 = f^0(\varepsilon_{33} + \varepsilon_{31} + \varepsilon_{32}) \tag{56-3}$$

$$\tilde{\sigma}^{ij} = \sigma^{kl}(F_k^i F_l^j - \delta_k^i \delta_l^j) \approx -2\varepsilon_{ij}\sigma^0, \quad i \neq j \tag{56-4}$$

Hence, the exterior deformation stress tensor is defined as:

$$\sigma_1^1 = f^0 \varepsilon_{11}, \quad \sigma_2^2 = f^0 \varepsilon_{22}, \quad \sigma_3^3 = f^0 \varepsilon_{33} \tag{57-1}$$

$$\sigma_j^i = f^0 \varepsilon_{ij} - 2\sigma^0 \varepsilon_{ij}, \quad i \neq j \tag{57-2}$$

As it is well-known, for isotropic simple materials, the classical stress-strain equation (constitutive equation) is:

$$\sigma_j^i (= \sigma_{ij}) = \lambda(\varepsilon_{ll})\delta_{ij} + 2\mu\varepsilon_{ij} \tag{58}$$

By comparing the shear components, it is clear that:

$$\mu = \frac{1}{2}(f^0 - 2\sigma^0). \tag{59}$$

As the first invariant of deformation stress is:

$$\sigma_1^1 + \sigma_2^2 + \sigma_3^3 = f^0 \Delta = (3\lambda + 2\mu)\Delta, \quad \Delta = \varepsilon_{11} + \varepsilon_{12} + \varepsilon_{13} \tag{60}$$

So, we have: $f^0 = 3\lambda + 2\mu = 3\lambda + f^0 - 2\sigma^0$. Then, we have:

$$\lambda = \frac{2\sigma^0}{3}. \tag{61}$$

However, from above formulation, it is found that, the following geometrical condition is implied:

$$F_j^i = F_i^j \tag{62}$$

This buried geometrical condition in classical constitutive equations is named as 'deformation conformal condition'. Many researches have been done on this topic in different formulation forms.

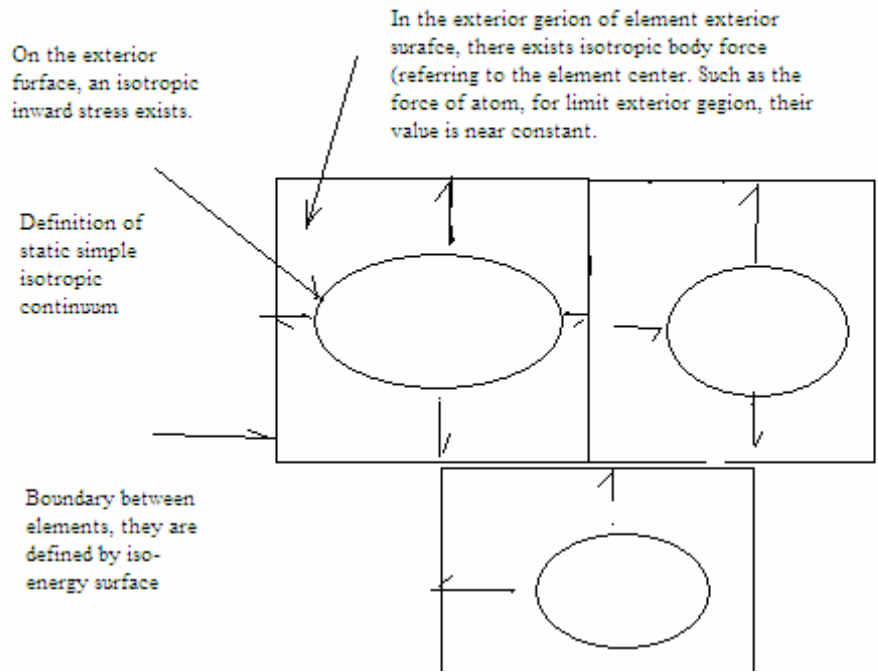

Figure 7. Isotropic Simple Solid Materials

Summering above analysis, the isotropic simple material is defined by intrinsic field:

$$f^1 = f^2 = f^3 = f^0, \quad \sigma^{ij} = \sigma^{ji} = -\sigma^0, i \neq j \tag{63-1}$$

Their Lame constants ($\lambda, \mu$) are determined by the intrinsic field as:

$$\lambda = \frac{2\sigma^0}{3}, \quad \mu = \frac{1}{2}(f^0 - 2\sigma^0) \tag{63-2}$$

Physically, the isotropic simple material has isotropic body force and isotropic exterior intrinsic stress both take the material element geometrical center as the start point of orientation direction. By Equation (44), the isotropic simple material has an intrinsic viscous stress:

$$\eta_{ij}^{vis} = -\frac{1}{2}(\sigma^{ij} + \sigma^{ji}) = \sigma^0 \tag{64}$$

Hence, the Lame constant $\lambda$ is adhesive force on material element. This interpretation is well known and explained in text books.

This simple case shows that the stress formulation in this research explains the material features in intrinsic sense.

As two special cases: *1) Incompressible Fluids:* for $\lambda = \frac{2\sigma^0}{3} \approx 0$, that is when the exterior surface adhesive (viscous) stress is near zero, the simple isotropic continuum is ideal fluid. Since on the material element surface the exterior intrinsic stress is isotropic, the micro fluid element can be thought as a sphere ball. The zero exterior surface force means that any closed surface can be treated as the material element surface, therefore, the fluid has no fixed configuration. By $\mu \approx \frac{1}{2} f^0$, the fluid viscous stress is produced by the exterior body force. *2) Discrete Medium:* for $f^0 = 2\sigma^0$, $\mu \approx 0$, the discrete medium is defined. For this kind of medium, $\sigma^{ij} = 0, i \neq j$ means that material elements only have contact relation such as sands (purely body force system, such as the ideal gas in classical thermodynamics).

### 5.2 *Polar Simple Fluid Material*

In classical infinitesimal deformation theory, the classical Stokes asymmetrical strain is defined as:

$$\omega_{ij} = \frac{1}{2}(F_j^i - F_i^j), \quad \varepsilon_{ij} = 0, \quad i \neq j \tag{65}$$

For polar simple fluid materials, defined as: $f^1 = f^2 = f^3 = \tilde{f}^0$, $\sigma^{ij} = -\sigma^{ji} = \tilde{\sigma}^0, i \neq j$. (For its static definition, referring Figure 7, but the isotropic stress in exterior surface is replaced by right-hand chirality rotational stress). Under this condition, for deformation $\vec{g}_i = F_i^j \vec{g}_j^0$, the exterior surface variation is:

$$\tilde{\sigma}^{ij} = \sigma^{kl}(F_k^i F_l^j - \delta_k^i \delta_l^j) \approx 2\omega_{ij}\sigma^0, \quad i \neq j \tag{66}$$

The exterior deformation stress field is:

$$\sigma_1^1 = f^0 \varepsilon_{11}, \quad \sigma_2^2 = f^0 \varepsilon_{22}, \quad \sigma_3^3 = f^0 \varepsilon_{33} \tag{67-1}$$

$$\sigma_j^i = -\sigma_i^j \approx 2\omega_{ij}\tilde{f}^0, \quad i \neq j \tag{67-2}$$

When the average isotropic stress $-p$ (fluid pressure) is defined as:

$$-p = \frac{1}{3}(f^0 \Delta), \Delta = \varepsilon_{11} + \varepsilon_{22} + \varepsilon_{33} \tag{68}$$

The constitutive equation is formulated as:

$$\sigma_j^i = -p\delta_{ij} + 2\tilde{\mu}\omega_{ij} \tag{69}$$

Where, $\tilde{\mu} = \frac{1}{2}\tilde{f}^0$ and the index is lowered to follow conventions. This equation is very common in fluid dynamics research.

Comparing with Equation (58) $\sigma_{ij} = \lambda(\varepsilon_{ll})\delta_{ij} + 2\mu\varepsilon_{ij}$, a more general form of constitutive equation for simple materials is:

$$\sigma_{ij}(=\sigma_j^i) = \lambda(\varepsilon_{ll})\delta_{ij} + 2\mu\varepsilon_{ij} + 2\tilde{\mu}\omega_{ij} \tag{70}$$

Where, $\lambda = \dfrac{2\sigma^0}{3}$, $\mu = \dfrac{1}{2}(f^0 - 2\sigma^0)$, $\tilde{\mu} = \dfrac{1}{2}\tilde{f}^0$. In this case, the solid-featured shear and fluid-featured shear are different. Their physical backgrounds are different. For elastic fluid concept, this equation is applicable.

This equation explains a long-standing question in my brain: why Stokes asymmetrical strain is discarded in solid deformation mechanics theory, while in fluid dynamics the Stokes asymmetrical strain plays the main role. (In classical solid mechanics, $\mu = \tilde{\mu}$ is used).

### 5.3 Linear Constitutive Equation in General Form

Strictly speaking, the classical strain definition omits the asymmetrical components of deformation tensor. Then, the deformation stress tensor is symmetrical. This shortage is well exposed in many researches. Here, a more exact linear formulation will be present.

In fact, the actual way is to define:

$$(F_i^k F_j^l - \delta_i^k \delta_j^l) = \frac{1}{2}(F_i^k - \delta_i^k)(F_j^l + \delta_j^l) + \frac{1}{2}(F_i^k + \delta_i^k)(F_j^l - \delta_j^l) \tag{71}$$

Hence,

$$\tilde{\sigma}^{kl} = \sigma^{ij}(F_i^k F_j^l - \delta_i^k \delta_j^l) = \frac{1}{2}\sigma^{ij}(F_j^l + \delta_j^l)(F_i^k - \delta_i^k) + \frac{1}{2}\sigma^{ij}(F_i^k + \delta_i^k)(F_j^l - \delta_j^l)$$

$$\approx \sigma^{il}(F_i^k - \delta_i^k) + \sigma^{kj}(F_j^l - \delta_j^l) \tag{72}$$

As,

$$\sigma^{il}(F_i^k - \delta_i^k) + \sigma^{kj}(F_j^l - \delta_j^l)$$
$$= \sigma^{mn}(F_i^j - \delta_i^j)\delta_n^l \delta_m^i \delta_j^k + \sigma^{nm}(F_i^j - \delta_i^j)\delta_n^k \delta_m^i \delta_j^l$$
$$= (\sigma^{mn}\delta_n^l \delta_m^i \delta_j^k + \sigma^{nm}\delta_n^k \delta_m^i \delta_j^l)(F_i^j - \delta_i^j) \tag{73}$$
$$= \tilde{C}_{lj}^{ki}(F_i^j - \delta_i^j)$$

Introducing parameters:

$$\tilde{C}_{lj}^{ki} = (\sigma^{mn}\delta_n^l \delta_m^i \delta_j^k + \sigma^{nm}\delta_n^k \delta_m^i \delta_j^l) \tag{74}$$

Then, interchange the index $i \leftrightarrow k$, $j \leftrightarrow l$, the exterior surface force variation is:

$$\tilde{\sigma}^{ij} = \tilde{C}_{jl}^{ik}(F_k^l - \delta_k^l), \quad i \neq j \tag{75}$$

Where, $\tilde{C}_{jl}^{ik} = (\sigma^{mn}\delta_n^j \delta_m^k \delta_l^i + \sigma^{nm}\delta_n^i \delta_m^k \delta_l^j)$.

Defining the material features parameter as:

$$C_{jl}^{ik} = \frac{1}{2}(\tilde{C}_{jl}^{ik} + \tilde{C}_{il}^{jk}) = \frac{1}{2}(\sigma^{mn} + \sigma^{nm})(\delta_n^j \delta_m^k \delta_l^i + \delta_n^i \delta_m^k \delta_l^j) \tag{76}$$

It shows that only symmetrical parts of exterior surface have contribution to deformation stress.

The exterior deformation stress fields are:

$$\sigma_1^1 = f^1(F_1^1 - 1), \quad \sigma_2^2 = f^2(F_2^2 - 1), \quad \sigma_3^3 = f^3(F_3^3 - 1) \tag{77-1}$$

$$\sigma_j^i = f^{(j)} F_j^i + C_{jl}^{ik}(F_k^l - \delta_k^l), \quad i \neq j \tag{77-2}$$

This is the ***generalized linear constitutive equation***.

The weak non-linear material parameters can be defined as:

$$C_{jl}^{ik}(\varepsilon) = \frac{1}{2}[\sigma^{mn}(F_n^l + \delta_n^l)\delta_j^k \delta_m^i + \sigma^{nm}(F_n^k + \delta_n^k)\delta_m^i \delta_j^l] \quad (78)$$

In this case, the material parameters have a weak dependence on deformation tensor.

The weak linear dependence can be expressed as the linear function of strain:

$$C_{jl}^{ik}(\varepsilon) = C_{jl}^{ik} + \frac{1}{2}[\sigma^{mn}\delta_j^k \delta_i^m \varepsilon_{nl} + \sigma^{nm}\delta_i^m \delta_j^l \varepsilon_{nk}] \quad (79)$$

The weak linear dependence relations are widely used in engineering mechanics for different reasoning and formulation. Here, the formulation can be viewed as a general form.

## 6. Stress Formulation for Chen S+R Additive Form Deformation

However, the isolated consideration of strain definition and deformation stress definition are not the best way. Here, a new way is explained.

From geometrical consideration, for general deformation $\vec{g}_i = F_i^j \vec{g}_j^0$, based on deformation geometry, the deformation tensor can be decomposed as the Chen Form-1:

$$F_j^i = S_j^i + R_j^i(\Theta) \quad (80)$$

Where, $S_j^i$ is symmetrical tensor, $R_j^i(\Theta)$ is an unit orthogonal rotation tensor with rotation angle $\Theta$.

Or Chen Form-2:

$$F_j^i = \tilde{S}_j^i + \frac{1}{\cos\theta}\tilde{R}_j^i(\theta) \quad (81)$$

Where, $\tilde{S}_j^i$ is symmetrical tensor, $\tilde{R}_j^i(\theta)$ is an unit orthogonal rotation tensor with rotation angle $\theta$.

Their deformation stress tensors will be discussed bellow, respectively.

### 6.1 Stress for Chen Form-1 Deformation

For Chen Form-1 deformation, $F_j^i = S_j^i + R_j^i$.

Case 1: If there is no local rotation, the deformation is: $F_j^i = S_j^i + \delta_j^i$, the deformation stress tensor is:

$$\sigma_1^1 = f^1 S_1^1, \quad \sigma_2^2 = f^2 S_2^2, \quad \sigma_3^3 = f^3 S_3^3 \quad (82\text{-}1)$$

$$\sigma_j^i = \frac{1}{2}(f^{(j)} + f^{(i)})S_i^j + C_{jl}^{ik} S_k^l, \quad i \neq j \quad (82\text{-}2)$$

Case 2: If there is no intrinsic stretching, the deformation is: $F_j^i = R_j^i$, as the exterior surface variation is: $\tilde{\sigma}^{ij} = \sigma^{kl}(F_k^i F_l^j - \delta_k^i \delta_l^j) = 0$ (this is because that the orthogonal rotation will not change the symmetry feature of $\sigma^{kl}$, while the anti-symmetrical parts have no contribution to deformation stress), and the unit orthogonal rotation tensor can be expressed as:

$$R_i^j = \delta_i^j + \sin\Theta \cdot L_i^j + (1-\cos\Theta)L_i^l L_l^j = \delta_i^j + \sin\Theta \cdot e_{ijk}L_k + (1-\cos\Theta)(L_i L_j - \delta_{ij}) \quad (83)$$

where $L_i$ is the components of unit rotation axe direction vector), the deformation stress tensor is:

$$\sigma_1^1 = f^1(1-\cos\Theta)(L_1 L_1 - 1), \tag{84-1}$$

$$\sigma_2^2 = f^2(1-\cos\Theta)(L_2 L_2 - 1), \tag{84-2}$$

$$\sigma_3^3 = f^3(1-\cos\Theta)(L_3 L_3 - 1) \tag{84-4}$$

$$\sigma_j^i = f^{(j)}(R_j^i - \delta_j^i), \quad i \neq j \tag{84-5}$$

Any Chen Form-1 deformation can be viewed as the stack of the two deformations. Hence, based on the **generalized linear constitutive equation, stress for Chen Form-1 Deformation is:**

$$\sigma_1^1 = f^1[S_1^1 + (1-\cos\Theta)(L_1 L_1 - 1)] \tag{85-1}$$

$$\sigma_2^2 = f^2[S_2^2 + (1-\cos\Theta)(L_2 L_2 - 1)] \tag{85-2}$$

$$\sigma_3^3 = f^3[S_3^3 + (1-\cos\Theta)(L_3 L_3 - 1)] \tag{85-3}$$

$$\sigma_j^i = \frac{1}{2}(f^{(j)} + f^{(i)})S_i^j + f^{(j)}(R_j^i - \delta_j^i) + C_{jl}^{ik} S_k^l, \quad i \neq j \tag{85-4}$$

This constitutive equation can be used for complicated materials. Here, it is the first time to give its intrinsic form.

**6.2 Stress for Chen Form-2 Deformation**

For Chen Form-2 deformation, $F_j^i = \tilde{S}_j^i + \frac{1}{\cos\theta}\tilde{R}_j^i(\theta)$, the orthogonal rotation is:

$$\begin{aligned}
\frac{1}{\cos\theta}\tilde{R}_j^i &= \delta_j^i + \frac{\sin\theta}{\cos\theta}\tilde{L}_j^i + (\frac{1}{\cos\theta} - 1)(\tilde{L}_k^i \tilde{L}_j^k + \delta_j^i) \\
&= \delta_j^i + \frac{\sin\theta}{\cos\theta} e_{ijk}\tilde{L}_k + (\frac{1}{\cos\theta} - 1)\tilde{L}_i \tilde{L}_j
\end{aligned} \tag{86}$$

For pure rotational deformation, $F_j^i = \frac{1}{\cos\theta}\tilde{R}_j^i(\theta)$, the surface forces variations are:

$$\tilde{\sigma}^{ij} = (\frac{1}{\cos^2\theta} - 1)\sigma^{ij} \tag{87}$$

Hence, for pure rotational deformation, the stress field is:

$$\sigma_1^1 = f^1(\frac{1}{\cos\theta} - 1)\tilde{L}_1 \tilde{L}_1 \tag{88-1}$$

$$\sigma_2^2 = f^2(\frac{1}{\cos\theta} - 1)\tilde{L}_2 \tilde{L}_2 \tag{88-2}$$

$$\sigma_3^3 = f^3(\frac{1}{\cos\theta} - 1)\tilde{L}_3 \tilde{L}_3 \tag{88-3}$$

$$\sigma_j^i = f^{(j)}(\frac{1}{\cos\theta}R_j^i - \delta_j^i) + \frac{1}{2}(\frac{1}{\cos^2\theta} - 1)(\sigma^{ij} + \sigma^{ji}), \quad i \neq j \tag{88-4}$$

By Equation (44), the item $\frac{1}{2}(\frac{1}{\cos^2\theta} - 1)(\sigma^{ij} + \sigma^{ji})$ is the adhesive (viscous) stress on exterior surface of material element. Hence, this Chen Form-2 deformation is mainly related with turbulence or fatigue-cracking.

Following the similar way as the last sub-section, based on the *generalized linear constitutive equation, stress for Chen Form-2 Deformation is:*

$$\sigma_1^1 = f^1[\tilde{S}_1^1 + (\frac{1}{\cos\theta} - 1)\tilde{L}_1\tilde{L}_1] \tag{88-1}$$

$$\sigma_2^2 = f^2[\tilde{S}_2^2 + (\frac{1}{\cos\theta} - 1)\tilde{L}_2\tilde{L}_2] \tag{88-2}$$

$$\sigma_3^3 = f^3[\tilde{S}_3^3 + (\frac{1}{\cos\theta} - 1)\tilde{L}_3\tilde{L}_3] \tag{88-3}$$

$$\sigma_j^i = \frac{1}{2}(f^{(j)} + f^{(i)})\tilde{S}_i^j + f^{(j)}(\frac{1}{\cos\theta}\tilde{R}_j^i - \delta_j^i) + \frac{1}{2}(\frac{1}{\cos^2\theta} - 1)(\sigma^{ij} + \sigma^{ji}) + C_{jl}^{ik}\tilde{S}_k^l, \quad i \neq j \tag{88-4}$$

This constitutive equation is different from Equation (85). For the same deformation, it seems to be that two different decompositions will produce different form of constitutive equation. This concept is wrong. In fact, both equations are derived from the general linear constitutive equations. But, these two constitutive equations do clear the different contribution of deferent deformation mechanisms. This is very important when the real engineering materials are concerned.

By Equation (44), the last Equation (88-4) can be written as:

$$\sigma_j^i = \frac{1}{2}(f^{(j)} + f^{(i)})\tilde{S}_i^j + f^{(j)}(\frac{1}{\cos\theta}\tilde{R}_j^i - \delta_j^i) - (\frac{1}{\cos^2\theta} - 1)\eta_{ij}^{vis} + C_{jl}^{ik}\tilde{S}_k^l, \quad i \neq j \tag{89}$$

Summing up above results, the constitutive equations based on Chen form decomposition of deformation tensor can be used to express the stress caused by local rotation in a much simple way.

## 7. Material Parameter Invariance Feature

Observing the static condition Equation (46) (for easy reading, they are rewritten at bellow) for a macro static continuum, it is found that the exterior original surface $\sigma^{ij}$ is determined by the interior region body force and the exterior region body forces.

$$\frac{1}{2}(\sigma^{ij} - \sigma^{ji}) + e^{ijk}f_{int}^k = 0, \text{ interior region} \tag{46-1}$$

$$\frac{\partial f_{ext}^{(i)}}{\partial x^i} = 0, \text{ exterior region} \tag{46-2}$$

$$\frac{1}{2}(\frac{\partial f_{ext}^j}{\partial x^i} + \frac{\partial f_{ext}^i}{\partial x^j}) + \frac{1}{2}(\sigma^{ij} + \sigma^{ji}) = 0, \quad i \neq j, \text{ on surface} \tag{46-3}$$

Observing the general stress definition Equation (52), (for easy reading, they are rewritten at bellow), it is found that when the deformation is removed ($F_j^i \Rightarrow \delta_j^i$) after an arbitral deformation, the stress becomes to zero.

$$\sigma_1^1 = f_{ext}^1 \cdot (F_1^1 - 1), \quad \sigma_2^2 = f_{ext}^2 \cdot (F_2^2 - 1), \quad \sigma_3^3 = f_{ext}^3 \cdot (F_3^3 - 1) \tag{52-1}$$

$$\sigma_j^i = f_{ext}^{(f)} F_j^i + \frac{1}{2}(\tilde{\sigma}^{ij} + \tilde{\sigma}^{ji}), \quad i \neq j \tag{52-2}$$

Where, $\tilde{\sigma}^{kl} = \sigma^{ij}(F_i^k F_j^l - \delta_i^k \delta_j^l)$, $k \neq l$, $i \neq j$.

However, by the stress definition, after the deformation process, the zero stress only gives out a conclusion: the viscous stress $\eta_{ij}^{vis} = -\frac{1}{2}(\tilde{\sigma}^{ij} + \tilde{\sigma}^{ji})$ is returned to its original value. Hence, possible variations on $\frac{1}{2}(\sigma^{ij} - \sigma^{ji})$ can be produced during the deformation process. This item will affect the internal region of material element. If the interior region body force is changed, in general sense, the material has changed (more or less). This means the evolution process of materials.

To describe the material evolution process, the interior region is studied bellow simply. The interior deformation stress principle components are defined by Equation (53) (for easy reading, they are rewritten at bellow) as:

$$\tilde{\sigma}_{int}^{i} = \frac{1}{2}e^{ijk}(\tilde{\sigma}^{jk} - \tilde{\sigma}^{kj}) = \frac{1}{2}e^{ijk}\sigma^{mn}(F_m^j F_n^k - F_m^k F_n^j) \tag{53}$$

Defining the 2-form of it as:

$$\tilde{\sigma}_{int} = \frac{1}{2}\tilde{\sigma}^{ij}\vec{g}_i^0 \wedge \vec{g}_j^0 \tag{90}$$

Then, one has the 3-form defined by the surface force variation (divergence):

$$d\tilde{\sigma}_{int} = \frac{1}{6}e^{ijk}\frac{\partial \tilde{\sigma}^{jk}}{\partial x^i}\vec{g}_i^0 \wedge \vec{g}_j^0 \wedge \vec{g}_k^0 \tag{91}$$

On physical sense, the 3-form is the energy passing through the exterior surface of material element. Therefore, an internal energy item $U = \omega \vec{g}_1 \wedge \vec{g}_2 \wedge \vec{g}_3$ can be introduced.

In initial 'static' state, defined by macro static continuum, the initial static internal energy is defined by:

$$U_0 = \omega_0 \vec{g}_1^0 \wedge \vec{g}_2^0 \wedge \vec{g}_3^0 \tag{92}$$

Where, by definition of macro static continuum, it is not zero:

$$\omega_0 = \frac{1}{3}\left(\frac{\partial f_{int}^1}{\partial x^1} + \frac{\partial f_{int}^2}{\partial x^2} + \frac{\partial f_{int}^3}{\partial x^3}\right) \tag{93}$$

After deformation, the current internal energy is defined by:

$$U = \frac{1}{3}\omega_0\sqrt{\frac{g}{g_0}}\vec{g}_1^0 \wedge \vec{g}_2^0 \wedge \vec{g}_3^0 \tag{94}$$

Where, the $\sqrt{\frac{g}{g_0}} \approx S_1^1 + S_2^2 + S_3^3$. Therefore, the internal energy variation is:

$$d\omega = \frac{1}{3}(\sqrt{\frac{g}{g_0}} - 1)\omega_0 \approx \frac{1}{3}(S_1^1 + S_2^2 + S_3^3)\omega_0 \tag{95}$$

On the other hand, in deformed state, the interior region new 'static' condition is:

$$\frac{1}{2}e^{ijk}(\tilde{\sigma}^{ij} - \tilde{\sigma}^{ji}) + \Delta f_{int}^k = 0 \text{, interior region} \tag{96}$$

Where, $\Delta f_{int}^k = (\tilde{f}_{int}^k - f_{int}^k)$. Hence, by Stokes equation, one has:

$$d\omega = \frac{1}{3}[\frac{\partial(\tilde{\sigma}^{23} - \tilde{\sigma}^{32})}{\partial x^1} + \frac{\partial(\tilde{\sigma}^{31} - \tilde{\sigma}^{13})}{\partial x^2} + \frac{\partial(\tilde{\sigma}^{12} - \tilde{\sigma}^{21})}{\partial x^3}] \quad (97)$$

This internal energy variation is not recoverable by remove deformations. In classical elastic-plastic mechanics theory, this item is named as plastic energy.

Based on above formulation, the material evolution equation can be defined as:

$$d\omega = \int_0^t [\frac{\partial(\tilde{\sigma}^{23} - \tilde{\sigma}^{32})}{\partial x^1} + \frac{\partial(\tilde{\sigma}^{31} - \tilde{\sigma}^{13})}{\partial x^2} + \frac{\partial(\tilde{\sigma}^{12} - \tilde{\sigma}^{21})}{\partial x^3}]dt = \int_0^t (S_1^1 + S_2^2 + S_3^3)\omega_0 dt \quad (98)$$

Where, the time parameter shows that the internal energy variation is determined by deformation history. In rational mechanics theory, this is named as memory effects.

By this formulation, the exterior surface force evolution equation is obtained as bellow:

$$\frac{\partial \Delta \sigma^{23}(t)}{\partial x^1} - \frac{\partial \Delta \sigma^{32}(t)}{\partial x^1} = \int_0^t S_1^1 \omega_0 dt \quad (99\text{-}1)$$

$$\frac{\partial \Delta \sigma^{31}(t)}{\partial x^2} - \frac{\partial \Delta \sigma^{13}(t)}{\partial x^2} = \int_0^t S_2^2 dt \quad (99\text{-}2)$$

$$\frac{\partial \Delta \sigma^{12}(t)}{\partial x^3} - \frac{\partial \Delta \sigma^{21}(t)}{\partial x^3} = \int_0^t S_3^3 dt \quad (99\text{-}3)$$

Where, $\Delta \sigma^{ij}(t) = \sigma^{ij}(t) - \sigma^{ij}(0)$. Further more, by the recovery parts, after remove the deformation, one has:

$$\sigma^{ij}(t) + \sigma^{ji}(t) = \sigma^{ij}(0) + \sigma^{ji}(0) \quad (100)$$

The final *exterior surface force evolution equation* is obtained as:

$$\sigma^{ij}(t) = \sigma^{ij}(0) + e^{ijk} \int_{unit} (\frac{1}{2}\int_0^t S_k^k \omega_0 dt)dx^k \text{ , no summation} \quad (101)$$

It only supplies the anti-symmetrical parts. Hence, the material parameters are invariant.

$$C_{jl}^{ik} = \frac{1}{2}(\sigma^{mn} + \sigma^{nm})(\delta_n^j \delta_m^k \delta_l^i + \delta_n^i \delta_m^k \delta_l^j) \quad (102)$$

That means that, the constitutive equations are invariant. It is clear that the plasticity of deformation is well explained by this fact.

## 8. Conclusion

In this research, the physical field on macro 'static' continuum is established firstly with the exterior differential tolls, then the deformation is introduced to get the variation of the physical field. Starting from this treatment, the stress concept is studied by the physical variation. The paper gives out the related stress definition exactly. As a logic consequence, the general constitutive equations are obtained, where the initial physical fields are the material parameters.

The advantages of such a kind of constitutive equation are that: it is directly usable for detailed analysis of various materials based on physical understanding about the materials [36-41].

Further, the constitutive equations for Chen Form-1 and Chen Form-2 deformations are obtained. By this way, the Chen rational mechanics theory is complicated as a united system.

It is sure there are many problems are waiting to be studied further, that will be expressed by other papers in near futures.